\pdfoutput=1

\documentclass[11pt]{article}

\usepackage[]{ACL2023}

\usepackage{times}
\usepackage{latexsym}

\usepackage[T1]{fontenc}

\usepackage[utf8]{inputenc}

\usepackage{microtype}

\usepackage{inconsolata}

\usepackage{times}
\usepackage{soul}
\usepackage{url}
\usepackage[utf8]{inputenc}
\usepackage{caption}
\usepackage{graphicx}
\usepackage{amsmath}
\usepackage{amsthm}
\usepackage{booktabs}
\usepackage{algorithm}
\usepackage[switch]{lineno}
\usepackage{amsfonts}
\usepackage{subfigure}
\usepackage{CJKutf8}

\usepackage{algpseudocode}
\usepackage{subfigure}

\newcommand\cn[1]{\begin{CJK}{UTF8}{gbsn}#1\end{CJK}}

\title{ERNIE-Music: Text-to-Waveform Music Generation with Diffusion Models}

\author{
Pengfei Zhu, Chao Pang, Yekun Chai, Lei Li, Shuohuan Wang, Yu Sun, Hao Tian, Hua Wu \\
Baidu Inc. \\
\texttt{\{zhupengfei03, pangchao04, chaiyekun, wangshuohuan, sunyu02\}@baidu.com} \\
}

\begin{document}
\maketitle

\begin{abstract}

In recent years, the burgeoning interest in diffusion models has led to significant advances in image and speech generation. Nevertheless, the direct synthesis of music waveforms from unrestricted textual prompts remains a relatively underexplored domain. In response to this lacuna, this paper introduces a pioneering contribution in the form of a text-to-waveform music generation model, underpinned by the utilization of diffusion models.
Our methodology hinges on the innovative incorporation of free-form textual prompts as conditional factors to guide the waveform generation process within the diffusion model framework. Addressing the challenge of limited text-music parallel data, we undertake the creation of a dataset by harnessing web resources, a task facilitated by weak supervision techniques.
Furthermore, a rigorous empirical inquiry is undertaken to contrast the efficacy of two distinct prompt formats for text conditioning, namely, music tags and unconstrained textual descriptions. The outcomes of this comparative analysis affirm the superior performance of our proposed model in terms of enhancing text-music relevance.
Finally, our work culminates in a demonstrative exhibition of the excellent capabilities of our model in text-to-music generation. We further demonstrate that our generated music in the waveform domain outperforms previous works by a large margin in terms of diversity, quality, and text-music relevance.\footnote{Generated cases are available at \url{https://reurl.cc/94W4yO}}
\end{abstract}

\section{Introduction}


Music, as a sophisticated and profound human art form, possesses a unique capacity to evoke emotions, alter moods, and tell compelling stories through its intricate interplay of harmony, melody, and rhythm. In recent years, the realm of music generation has garnered significant attention and interest, coinciding with the rapid advancements in deep learning techniques.

Within this context, some research endeavors, exemplified by works such as~\cite{DBLP:journals/corr/abs-2211-11216}, have concentrated on the domain of symbolic music generation. This approach entails the acquisition of knowledge to predict sequences of musical composition, encompassing elements such as notes, pitch, and dynamic attributes. However, it is noteworthy that the resultant symbolic music lacks performance attributes, necessitating subsequent post-processing to synthesize the auditory experience of the musical piece. Conversely, an alternative line of inquiry, exemplified by works such as~\cite{DBLP:journals/corr/abs-2208-08706}, has been dedicated to the generation of audio or waveform-based music. Notably, this approach obviates the need for additional synthesis steps, as it directly produces audio signals. Nevertheless, it is important to recognize that generating audio signals in this manner often presents inherent challenges in controlling and fine-tuning performance attributes to achieve the desired level of quality and satisfaction.


Besides works on unconditional music generation, there have been explorations about conditional music generation~\cite{DBLP:journals/corr/abs-2208-08706,DBLP:journals/corr/abs-2211-11248}, which aims to meet the application requirements in scenarios such as automatic video soundtrack creation and music creation with specific genres or features. Notably, generative models can leverage information from various modalities, such as text and image, to create relevant outputs for a conditional generation. 

In addition to unconditional music generation, there is a growing interest in conditional music generation~\cite{DBLP:journals/corr/abs-2208-08706,DBLP:journals/corr/abs-2211-11248}. This field caters to specific application needs, like creating video soundtracks or generating music with specific genres or features. Generative models can use various data modalities, such as text and images, to create relevant outputs in conditional music generation. Nonetheless, the challenge of directly generating musical waveforms from unrestricted textual input remains a relatively underexplored frontier. While research efforts have delved into text-conditioned music generation, exemplified by works such as~\cite{DBLP:journals/corr/abs-2211-11216}, BUTTER~\cite{zhang2020butter}, and Mubert\footnote{https://github.com/MubertAI/Mubert-Text-to-Music}, it is noteworthy that these approaches do not possess the capability to directly produce musical audio based on unstructured free-form text prompts.

To address prior limitations, we introduce ERNIE-Music, a pioneering effort in free-form text-to-music generation using diffusion models in the waveform domain. To overcome the shortage of parallel text-to-music data, we have undertaken the collection of music waveforms along with their corresponding top-voted comments from the internet.  We employ conditional diffusion models to generate musical waveforms and investigate the impact of text format on enhancing text-music relevance.

To conclude, the contributions of this paper are:

\begin{itemize}

\item We introduce a music generation model that leverages free-form text as a conditioning factor, utilizing the diffusion model to generate waveform-based music.
\item We curate a dataset of free-form text-music parallel data from the internet.
\item We investigate and compare the efficacy of two text formats for conditioning the generative model, demonstrating that the use of free-form text significantly enhances text-music relevance.
\item Our results highlight the model's ability to produce diverse, high-quality music with markedly improved text-music relevance, surpassing existing methods by a large margin.
\end{itemize}
\section{Related Work}

\paragraph{Controllable Music Generation}


Controlled music generation faces the persistent challenge of effectively imposing constraints on musical output. Previous approaches have employed various techniques to address this issue. For instance, VQ-CPC~\cite{hadjeres2020vector} focuses on learning local music features devoid of temporal information. Meanwhile, \cite{DBLP:journals/corr/abs-2208-08706} leverages tempo information as a condition for generating music in the ``techno'' genre. BUTTER~\cite{zhang2020butter} adopts a natural language representation encompassing attributes like music key, meter, and style to exercise control over music generation. Furthermore, \cite{DBLP:journals/corr/abs-2211-11216} extends this exploration by investigating the impact of different pre-trained models in text-to-music generation.
Besides, retrieval-based methods can be adopted to generate music by combining human-created music pieces. Mubert firstly employs Sentence-BERT~\cite{reimers2019sentence} to encode input natural language and music tags, secondly retrieves relevant tags based on the distance among the representation, finally combines relevant sounds (all crafted by musicians and sound designers) to obtain the generated music.

\paragraph{Diffusion Models}

Diffusion models~\cite{DBLP:conf/icml/Sohl-DicksteinW15,DBLP:conf/nips/HoJA20} are latent variable models rooted in non-equilibrium thermodynamics. They operate by gradually disassembling the structure of the original data distribution through a progressive forward diffusion process and subsequently acquire the means to reconstruct the original data via a finite iterative denoising process.  In recent years, diffusion models have gained significant traction across diverse domains, including image generation~\cite{DBLP:conf/icml/NicholDRSMMSC22,DBLP:conf/nips/DhariwalN21,ramesh2022hierarchical} and audio generation~\cite{DBLP:conf/iclr/ChenZZWNC21,Kreuk2022AudioGenTG}. Our work is closely aligned with the realm of diffusion-based approaches in text-to-audio generation~\cite{DBLP:conf/iclr/ChenZZWNC21,Kreuk2022AudioGenTG}. Some concurrent works~\cite{DBLP:journals/corr/abs-2302-03917,DBLP:journals/corr/abs-2301-11757,DBLP:journals/corr/abs-2301-11325} use diffusion models to tackle the text-to-music generation, which mainly focus on given specific music genres or instruments. It is important to note that while these prior works primarily focus on speech generation or some restricted text descriptions, our research extends the application of diffusion models to the synthesis of music waveforms based on arbitrary textual prompts, representing a distinct task within the audio generation domain.



\section{Method}

This section commences with an overview of diffusion models, providing the overall context for our subsequent discussion. Subsequently, we delve into the specifics of our text-conditional diffusion model, elucidating its architecture and the objectives underpinning its training.

\begin{figure*}[!htb]
	\centering
	\includegraphics[width=0.82 \textwidth]{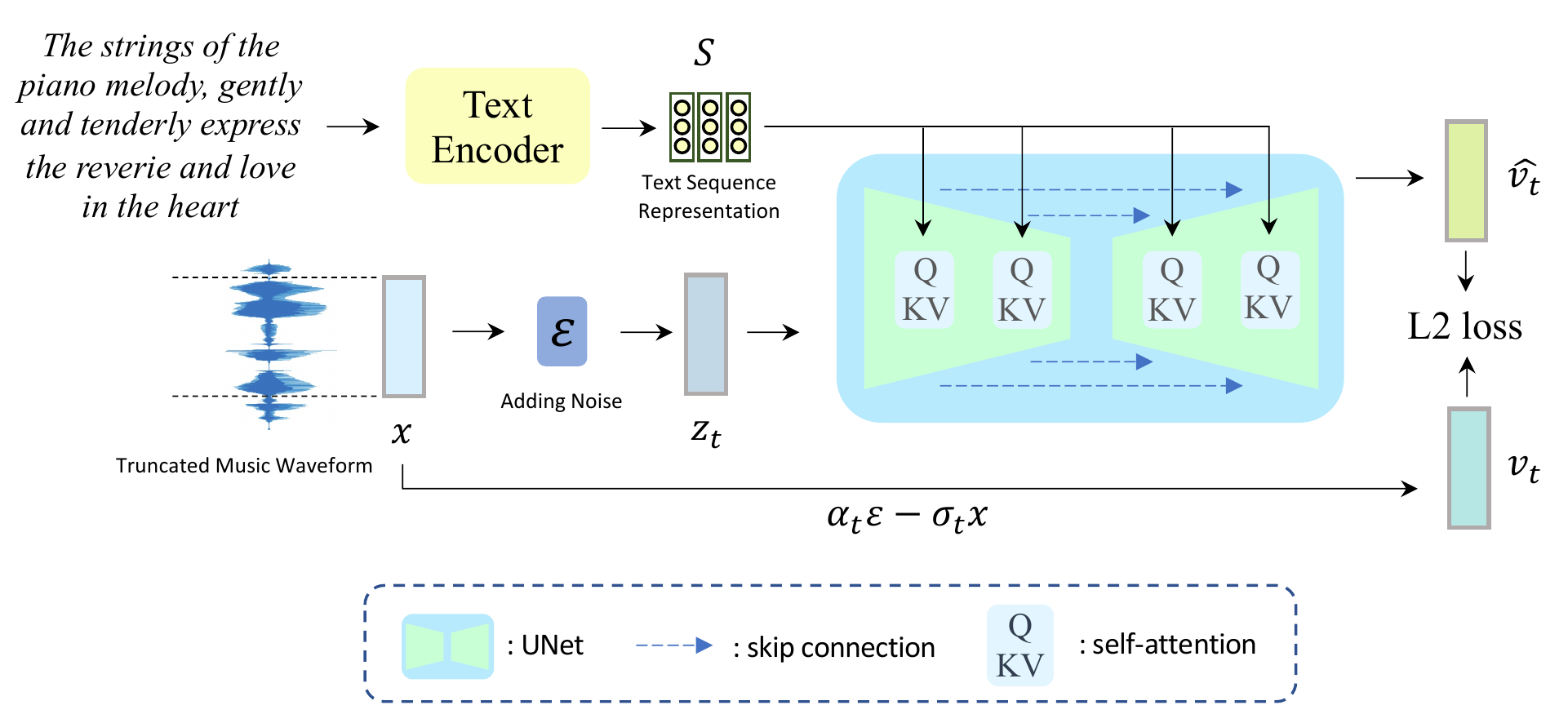}
	\caption{The overall architecture of text-to-music generation training. The text is input to the text encoder to obtain the sequence representation $S$, then $S$ and the sampled music waveform (noise added) $z_{t}$ are input to the UNet to obtain the estimated volocity $\hat{v}_t$, finally we calculate the L2 loss between $\hat{v}_t$ and the real volocity $v_t$. For the input text, the original Chinese is ``{\cn{钢琴旋律的弦音，轻轻地、温柔地倾诉心中的遐想、心中的爱恋}}''.}
	\label{fig:model}
\end{figure*}
\subsection{Unconditional Diffusion Model}


Diffusion models~\cite{DBLP:conf/icml/Sohl-DicksteinW15,DBLP:conf/nips/HoJA20} are composed of two essential components: a \textit{forward process}, where noise is progressively incorporated into a data sample, and a \textit{reverse process}, which subsequently removes this noise through multiple iterations to produce a sample that aligns with the authentic data distribution. Specifically, our approach is founded upon the diffusion model formulated within a continuous-time framework~\cite{DBLP:journals/corr/abs-2107-00630,DBLP:journals/corr/abs-1905-09883,DBLP:conf/iclr/ChenZZWNC21,DBLP:conf/iclr/0011SKKEP21,DBLP:conf/iclr/SalimansH22}.


In the context of diffusion models, we begin with a data sample denoted as $x$ drawn from the distribution $p(x)$. These models make use of latent variables $z_t$, where the parameter $t$ spans the continuous interval from $0$ to $1$. The log signal-to-noise ratio, represented as $\lambda_t$, is precisely defined as $\lambda_t = \log\left(\frac{\alpha_t^2}{\sigma_t^2}\right)$, where $\alpha_t$ and $\sigma_t$ correspond to the components of the noise schedule.


During the \textit{forward process}, often referred to as the \textit{diffusion process}, we progressively incorporate Gaussian noise into the sample, conforming to a Markov chain, characterized by the following progression:
\begin{align}
q(z_t|x)&=\mathcal N (z_t;\alpha_t x, \sigma_t^2 \textbf{I})  \\
q(z_{t'}|z_t) &= \mathcal N (z_{t'};(\alpha_{t'}/\alpha_t)z_t, \sigma_{t'|t}^2 \textbf{I}) 
\end{align}
where $t,t' \in [0,1]$ and $t<t'$, and $\sigma_{t'|t}^2=(1-e^{\lambda_{t'}-\lambda_t})\sigma_{t'}^2$.

In the \textit{reverse process}, a function approximation with parameters $\theta$ (denoted as $\hat{x}_\theta(z_t, \lambda_t, t) \approx x$) estimates the denoising procedure:
\begin{equation}
    p_\theta(z_t|z_{t'}) = \mathcal{N} (z_t; \tilde{\mu}_{t|t'}(z_{t'}, x)), \tilde{\sigma}_{t|t'}^2\textbf{I}) 
\end{equation}
where $\tilde{\mu}_{t|t'}(z_{t'}, x, t')) = e^{\lambda_{t'}-\lambda_t}(\alpha_t/\alpha_{t'})z_{t'} + (1-e^{\lambda_{t'}-\lambda_t})\alpha_t x$. 


Initiating from $z_1 \sim \mathcal{N}(0,\textbf{I})$, the \textit{reverse process} involves the sequential application of the denoising procedure to the latent variables $z_t$, ultimately yielding $z_0 = \hat{x}$. To train the denoising model $\hat{x}_\theta(z_t, \lambda_t, t)$, we adopt the weighted mean squared error loss as our optimization objective:

\begin{equation}\label{eqation_loss}
    L = E_{t \sim [0,1], \epsilon \sim \mathcal{N} (0,\textbf{I})}[w(\lambda_t) \lVert \hat{x}_\theta(z_t, \lambda_t, t)-x \rVert _2^2]
\end{equation}
where $w(\lambda_t)$ denotes the weighting function and $\epsilon \sim \mathcal{N} (0,\textbf{I})$ denotes the noise.

\subsection{Conditional Diffusion Model}

Numerous studies have effectively employed generative models within conditional settings~\cite{DBLP:journals/corr/MirzaO14,DBLP:conf/nips/SohnLY15,DBLP:conf/cvpr/RombachBLEO22}. In the case of conditional diffusion models, the focus shifts from approximating the distribution $p(x)$ to $p(x|y)$, achieved by modeling the denoising process as $\hat{x}_\theta(z_t, \lambda_t, t, y)$, where $y$ represents the conditioning variable. This conditioning variable $y$ can assume various modalities, encompassing image, text, and audio.

In the text-to-music generation scenario, $y$ takes the form of a textual prompt. This textual input serves as a guiding element for the model, steering it toward the generation of music that corresponds to the provided text. In subsequent sections, we delve into the intricacies of modeling the conditional diffusion model in detail.

\subsubsection{Model Architecture}\label{timestep_condition}


For text-to-music generation, our diffusion process conditions on textual input denoted as $y$. As illustrated in Figure \ref{fig:model}, our comprehensive model architecture comprises a conditional music diffusion model responsible for modeling the anticipated \textit{velocity}, represented as $\hat{v}_\theta(z_t, t, y)$~\cite{DBLP:conf/iclr/SalimansH22}. Additionally, we incorporate a text encoder denoted as ${\rm E}(\cdot)$, which transforms text tokens with a length of $n$ into a sequence of vector representations $[ s_0; S] $, each possessing a dimensionality of $d_E$. Here, $S=[ s_1, ..., s_n] $, with $s_i \in \mathbb{R}^{d_E}$, and $s_0$ serving as the classification representation of the input text.

The inputs to the music diffusion model encompass the latent variable $z_t \in \mathbb{R}^{d_c \times d_s}$, the timestep $t$ (which is subsequently transformed into the embedding $e_t \in \mathbb{R}^{d_t \times d_s}$), and the representation of the text sequence $[ s_0; S]  \in \mathbb{R}^{(n+1) \times d_E}$. Here, $d_c$ corresponds to the number of channels, $d_s$ signifies the sample size, and $d_t$ denotes the feature size of the timestep embedding. The output of this architecture is represented by the estimated \textit{velocity}, denoted as $\hat{v}_t \in \mathbb{R}^{d_c \times d_s} $.


Inspired by previous works on latent diffusion models~\cite{DBLP:conf/icml/NicholDRSMMSC22,DBLP:conf/cvpr/RombachBLEO22,DBLP:conf/nips/DhariwalN21}, we have adopted  the architecture of UNet~\cite{DBLP:conf/miccai/RonnebergerFB15} whose key components are stacked convolutional blocks and self-attention blocks~\cite{DBLP:conf/nips/VaswaniSPUJGKP17}. Generation models can estimate the conditional distribution, notably $p(x|y)$, and there exist various techniques to integrate conditional information $y$ into generative models~\cite{DBLP:conf/nips/SohnLY15}.

Our diffusion network is designed to predict the latent velocity, denoted as $\hat{v}_\theta$, at randomly sampled timestep $t$, leveraging the noised latent $z_t$ and a textual input $[ s_0; S]$ as conditioning elements. To integrate the conditioning information into the diffusion process, we employ a fusion operation, denoted as ${\rm Fuse}(\cdot, \cdot)$, on the timestep embedding $e_t$ and the text classification representation $s_0$. This operation yields a text-aware timestep embedding, ${e'}t = {\rm Fuse}(e_t, s_0) \in \mathbb{R}^{d{t'} \times d_s}$. Subsequently, we concatenate this modified embedding with $z_t$ to derive $z'_t = (z_t \oplus {e'}t) \in \mathbb{R} ^{(d{t'}+d_c)\times d_s}$. It is worth noting that, for simplicity, the operations involving the timestep embedding have been omitted from Figure~\ref{fig:model}. In Section \ref{compare_fuse}, we delve into a comparative analysis of different implementations of the fusion operation, ${\rm Fuse}(\cdot, \cdot)$, to evaluate their performance.

Furthermore, we incorporate the conditional representation into the self-attention blocks~\cite{DBLP:conf/nips/VaswaniSPUJGKP17}, which are responsible for capturing global information within the music signals. Within the self-attention blocks, taking into account the intermediate representation, where $z'_t \in \mathbb{R} ^{(d_t+d_c)\times d_s}$ is denoted as $\phi(z'_t) \in \mathbb{R}^{d_a \times d_\phi}$, and $S \in \mathbb{R}^{n \times d_E}$, the output is calculated in the following manner:


\begin{equation}
	{\rm Attention}(Q,K,V)={\rm softmax}(\frac{QK^T}{\sqrt{d_k}})V 
\end{equation}
\begin{equation}
 	{\rm head_i} = {\rm Attention}(Q_i, K_i, V_i) 
\end{equation}
\begin{equation}
    Q_i = \phi(z'_t) \cdot W_i^Q 
\end{equation}
\begin{equation}
    K_i = {\rm Concat}(\phi(z'_t) \cdot W_i^K,\, S \cdot W_i^{SK}) 
\end{equation}
\begin{equation}
    V_i = {\rm Concat}(\phi(z'_t) \cdot W_i^V,\, S \cdot W_i^{SV})  
\end{equation}
\begin{equation}
	{\rm CSA}(\phi(z'_t),\,S)={\rm Concat}({\rm head_1}, ..., {\rm head_h})W^O 
\end{equation}

\noindent where $W_i^Q \in \mathbb{R}^{d_{\phi} \times d_q}$, $W_i^K \in \mathbb{R}^{d_{\phi} \times d_k}$, $W_i^V \in \mathbb{R}^{d_{\phi} \times d_v}$, $W_i^{SK} \in \mathbb{R}^{d_{E} \times d_k}$, $W_i^{SV} \in \mathbb{R}^{d_{E} \times d_v}$, $W^O \in \mathbb{R}^{hd_{v} \times d_{\phi}}$ are parameter matrices, and $h$ denotes the number of heads, and ${\rm CSA}(\cdot, \cdot)$ denotes the conditional self-attention operation.

\subsubsection{Training}
Following~\cite{DBLP:conf/iclr/SalimansH22}, we set the weighting function in equation~\ref{eqation_loss} as the ``SNR+1'' weighting for a more stable denoising process. 

Specifically, for the noise schedule $\alpha_t$ and $\sigma_t$, we adopt the cosine schedule~\cite{DBLP:conf/icml/NicholD21} $\alpha_t={\rm cos}(\pi t / 2)$, $\sigma_t={\rm sin}(\pi t / 2)$, and the variance-preserving diffusion process satisfies $\alpha_t^2 + \sigma_t^2 = 1$. We denote the function approximation as $\hat{v}_\theta(z_t, t, y)$, where $y$ denotes the condition. The prediction target of $\hat{v}_\theta(z_t, t, y)$ is \textit{velocity} $v_t \equiv \alpha_t \epsilon - \sigma_t x$, which gives $\hat{x}=\alpha_tz_t-\sigma_t \hat{v}_\theta(z_t, t, y)$. Finally, our training objective is:
\begin{align}
L_\theta &=  (1+\alpha_t^2/\sigma_t^2) \lVert x-\hat{x}_t \rVert_2^2  \\
&= \lVert v_t - \hat{v}_t \rVert _2^2  
\end{align}

Algorithm \ref{alg:train} (in Appendix) displays the complete training process with the diffusion objective proposed by \cite{DBLP:conf/iclr/SalimansH22}.


\section{Experiments}

\begin{table}[!t]\small
\renewcommand\arraystretch{1.2}
	\centering
	\begin{tabular}{p{3.8cm}|p{1.2cm}|p{1.2cm}}
        \hline
         & Train & Test \\
		\hline
        Num. of Data Samples & 3890 & 204 \\
        \hline
        Avg. Text (Tokens) Length & 63.23 & 64.45 \\
        \hline
        Music Sample Rate & \multicolumn{2}{l}{16000} \\
        \hline
        Music Sample Size & \multicolumn{2}{l}{327680} \\
        \hline
        Music Duration & \multicolumn{2}{l}{20 seconds} \\
        \hline
	\end{tabular}
     \caption{\label{tab:data_statistics} The statistics of our collected dataset.}
\end{table}

\subsection{Implementation Details}
Following previous works~\cite{DBLP:conf/cvpr/RombachBLEO22,DBLP:conf/icml/NicholDRSMMSC22,DBLP:conf/nips/HoJA20}, we use UNet~\cite{DBLP:conf/miccai/RonnebergerFB15} architecture for the diffusion model. The UNet model uses 14 layers of stacked convolutional blocks and attention blocks for the downsample and upsample module, with skipped connections between layers with the same hidden size. It uses the input/output channels of 512 for the first ten layers and two 256s and 128s afterward. We employ the attention at 16x16, 8x8, and 4x4 resolutions. The sample size and sample rate of the waveform are 327,680 and 16,000, and the channel size is 2. The timestep embedding layer contains trainable parameters of 8x1 shape. It concatenates the noise schedule to obtain the embedding, which is then expanded to the sample size to obtain $e_t \in \mathbb{R}^{16 \times 327,680} $. For the text encoder ${\rm E}(\cdot)$, we use ERNIE-M~\cite{ouyang-etal-2021-ernie} to encode multi-lingual text inputs such as Chinese, English, Korean and Japanese, etc.
\subsection{Dataset}


Users on music platforms comments on music they like and they upvotes the comments they favor. We observe that popular comments with high upvotes is of high quality and contain useful music-related information such as musical instruments, genres, and human moods. Thus we collect a large set of text-music pairs data as training set.


The statistics of our collected Web Music with Text dataset and examples are listed in Table~\ref{tab:data_statistics} and~\ref{tab:example_data}. Note that the time duration of our collected music samples are usually 2 to 3 minutes, thus the actual number of training samples may be considered as 6 to 9 times larger than the music samples because of randomly sampling 20 seconds during the training process.


%
%
%

\subsection{Evaluation Metric}
For the text-to-music generation task, we evaluate performance in two aspects: text-music relevance and music quality. 
Because there is currently a lack of well-recognized and authoritative objective evaluation methods for text-music relevance, and the objective metrics for evaluation music quality such as Frechet Audio Distance (FAD) only calculate the closeness between the generated music and the real music instead of the actual quality~\cite{DBLP:journals/corr/abs-2211-11248}, we employ human evaluation methods. 
We use the compared methods or models to generate music based on texts from the test set, and manually score the generated music and calculate the mean score by averaging over results from different evaluators. We hire 10 people (who are of average listener annotator level among human) to perform human evaluation, scoring the music generated by each compared model, and then average the scores over the 10 people for each generated music. The identification of models corresponding to the generated music is invisible to the evaluators. Finally, we average the scores of the same model on the entire test samples to obtain the final evaluation results of the models.

\subsection{Compared Methods}

The methods for comparison are Text-to-Symbolic Music (denoted as TSM)~\cite{DBLP:journals/corr/abs-2211-11216}, Mubert and Musika~\cite{DBLP:journals/corr/abs-2208-08706}. The generated music from Mubert is actually created by human musicians, and TSM only generates music score, which needs to be synthesized into music audio by additional tools, so the music quality among Mubert, TSM, and our model is not comparable. Thus, we only compare the text-music relevance between them and our model. To synthesize the music audio based on the symbolic music score generated by TSM, we first adopt abcMIDI\footnote{https://github.com/sshlien/abcmidi} to convert the abc file output by TSM to MIDI file and then use FluidSynth\footnote{https://github.com/FluidSynth/fluidsynth} to synthesize the final music audio. For music quality, we compare our model's performance with Musika, a recent famous work that also directly generates waveform music.


\begin{table}[h]\small
\renewcommand\arraystretch{1.2}
	\centering
	\begin{tabular}{p{3cm}|p{1cm}|p{1cm}|p{1cm}}
		\hline
		Method & Score$\uparrow$ & Top Rate $\uparrow$ & Bottom Rate$\downarrow$ \\
        \hline
        TSM~\cite{DBLP:journals/corr/abs-2211-11216} & 2.05  & 12\% & 27\% \\ 
        \hline
        Mubert & 1.85 & 37\% & 32\%\\
        \hline
        our model & \textbf{2.43} & \textbf{55\%}  & \textbf{12\%}\\
		\hline
	\end{tabular}
 \caption{\label{tab:main_results} Comparison of text-music relevance.}
 \vspace{-0.5cm}
\end{table}

\begin{table}[h]\small
\renewcommand\arraystretch{1.2}
	\centering
	\begin{tabular}{p{3cm}|p{1cm}|p{1cm}|p{1cm}}
		\hline
		Method & Score$\uparrow$ & Top Rate$\uparrow$ & Bottom Rate$\downarrow$ \\
        \hline
        Musika~\cite{DBLP:journals/corr/abs-2208-08706} & 3.03 & 5\% & 13\% \\ 
        \hline
        our model & \textbf{3.63} & \textbf{15}\% & \textbf{2}\% \\
		\hline
	\end{tabular}
 \caption{\label{tab:quality_results} Comparison of music quality.}
\end{table}

\subsection{Results}

Table~\ref{tab:main_results} and \ref{tab:quality_results} show the evaluation results of text-music relevance and music quality. For text-music relevance evaluation, we use a ranking score of 3 (best), 2, 1 to denote which of the three models has the best relevance given a piece of text. For music quality, we use a five-level score of 5 (best), 4, 3, 2, 1, which indicates to what extent the evaluator prefers the melody and coherence of the music. The top rate means the probability that the music obtains the highest score, and the bottom rate means the probability that the music obtains the lowest score. The results indicate that our model can generate music with better quality and text-music relevance which outperforms related works by a large margin. 


\begin{figure*}[h]\small\centering
    \subfigure[]{
		\begin{minipage}[]{0.33\linewidth}
			\includegraphics[width=1\textwidth]{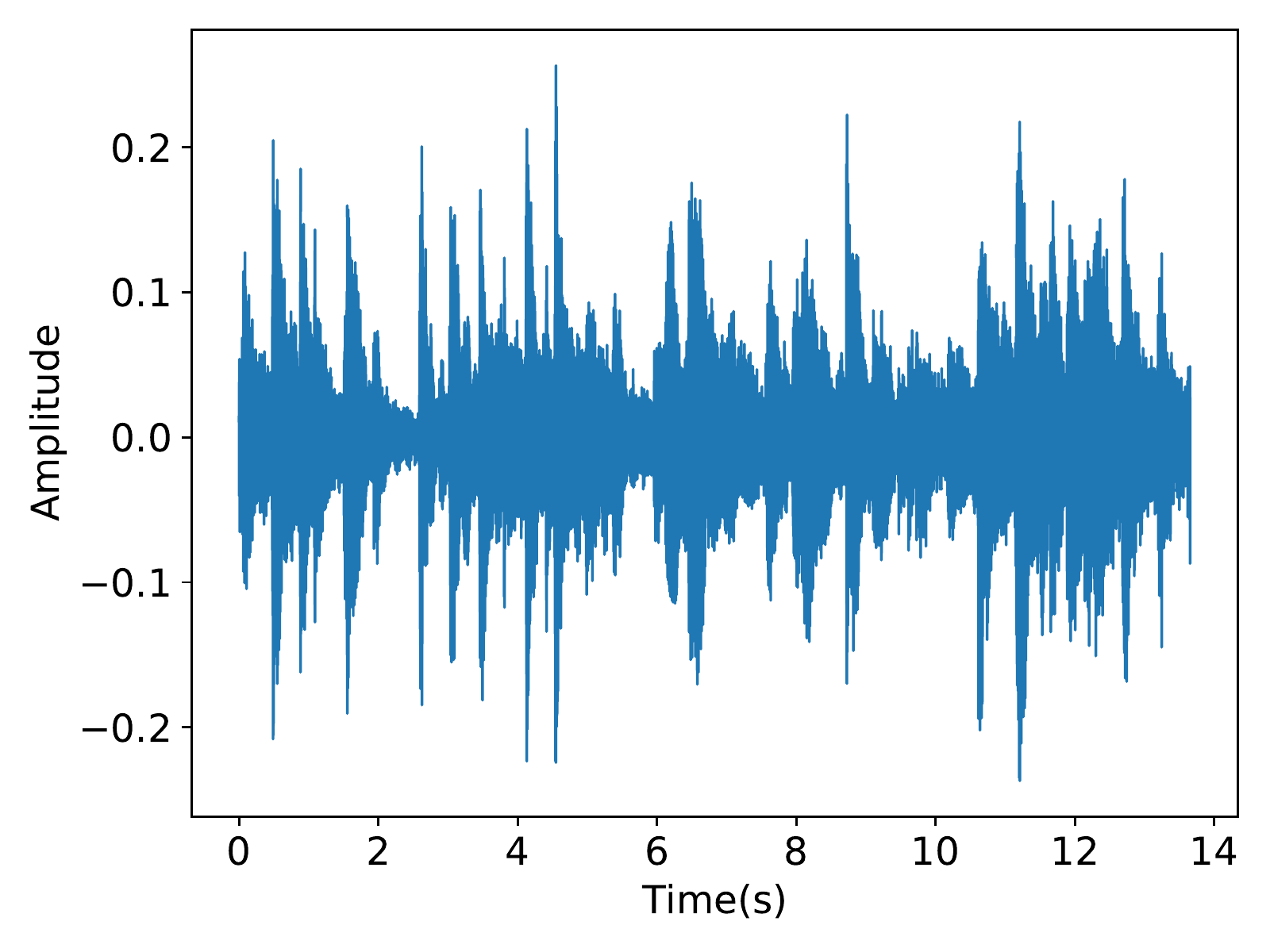}
		\end{minipage}
	}
	\subfigure[]{
		\begin{minipage}[]{0.33\linewidth}
			\includegraphics[width=1\textwidth]{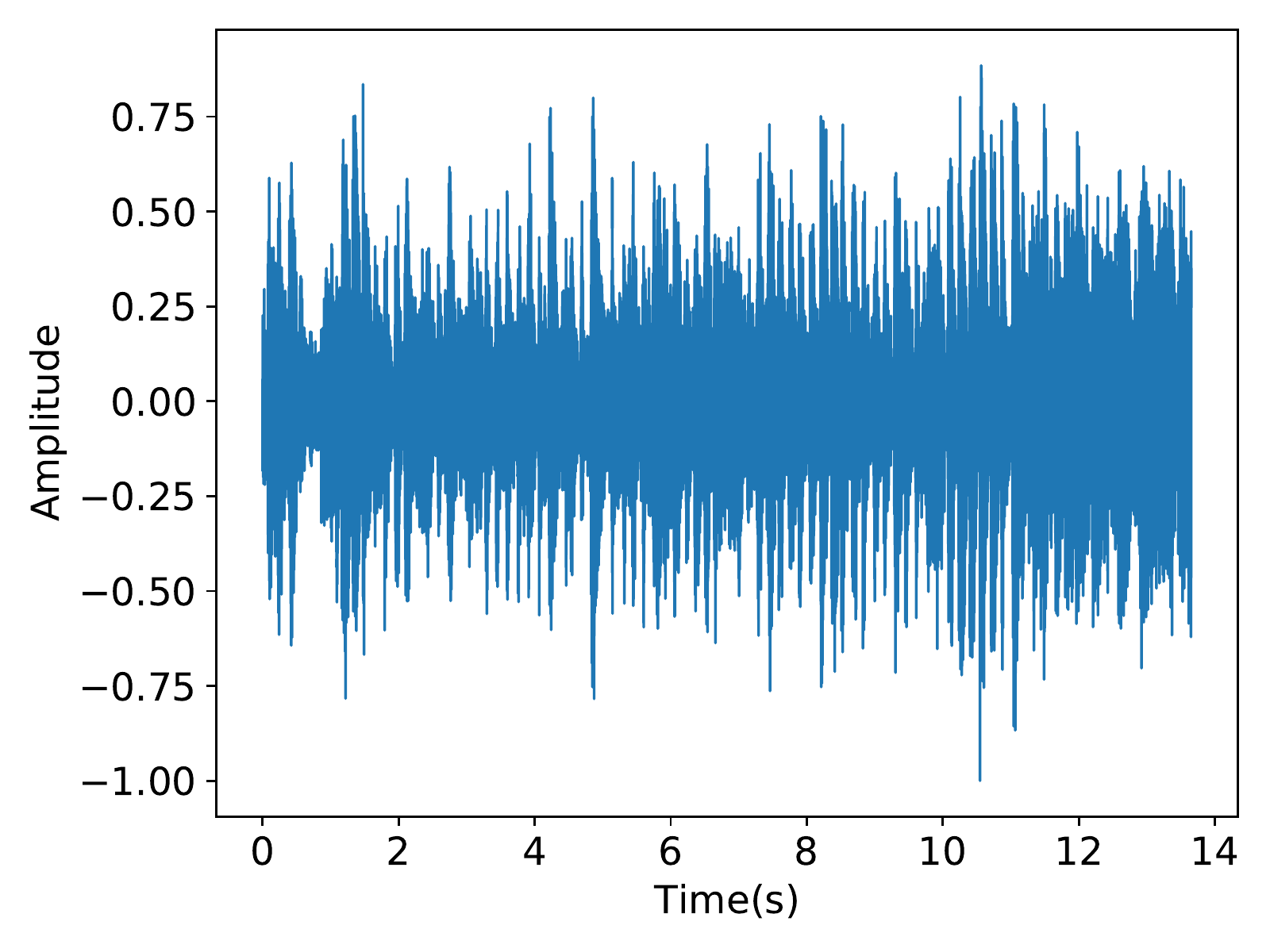}
		\end{minipage}
	}
	\subfigure[]{
		\begin{minipage}[]{0.33\linewidth}
			\includegraphics[width=1\textwidth]{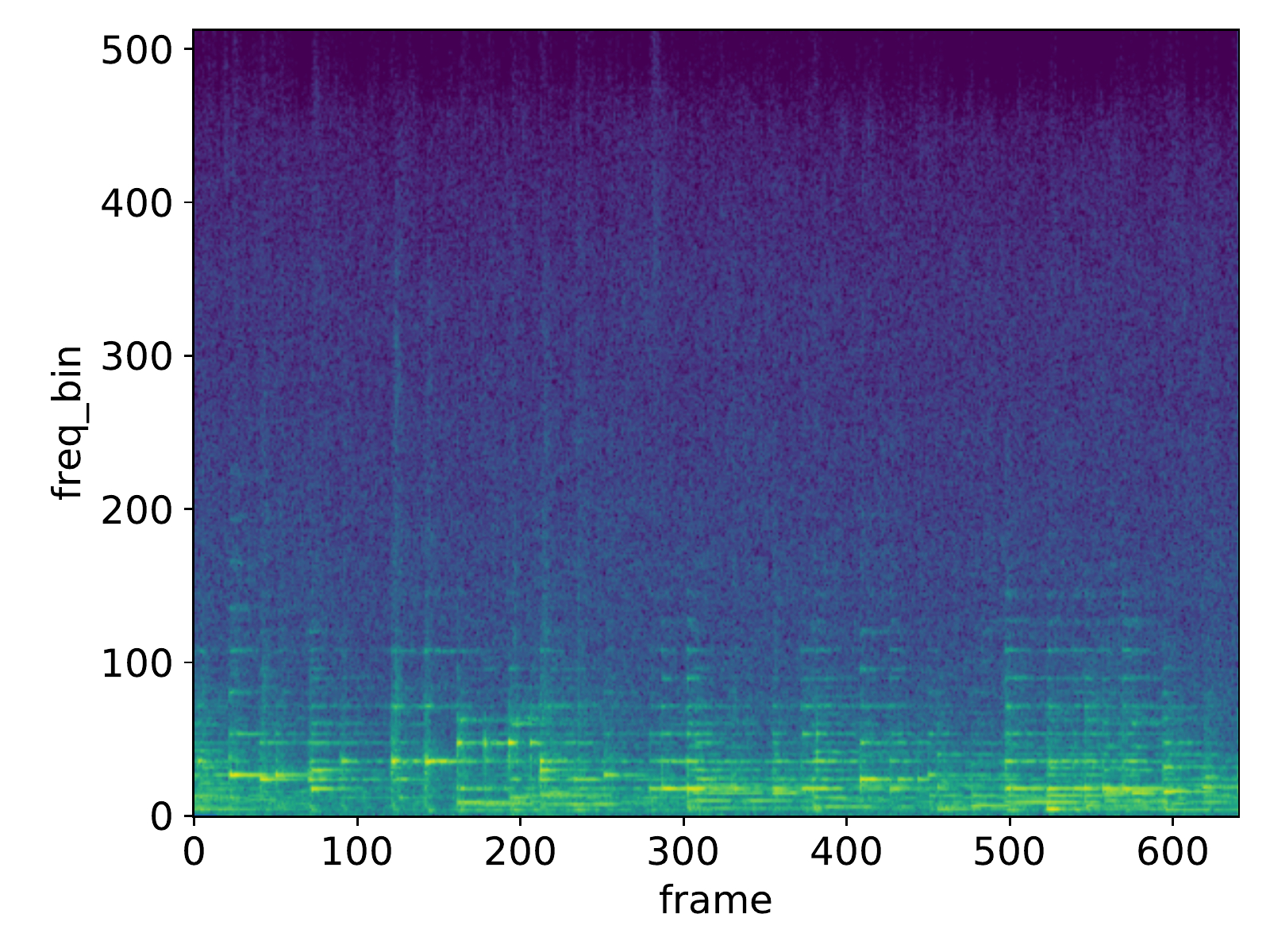}
		\end{minipage}
	}
	\subfigure[]{
		\begin{minipage}[]{0.33\linewidth}
			\includegraphics[width=1\textwidth]{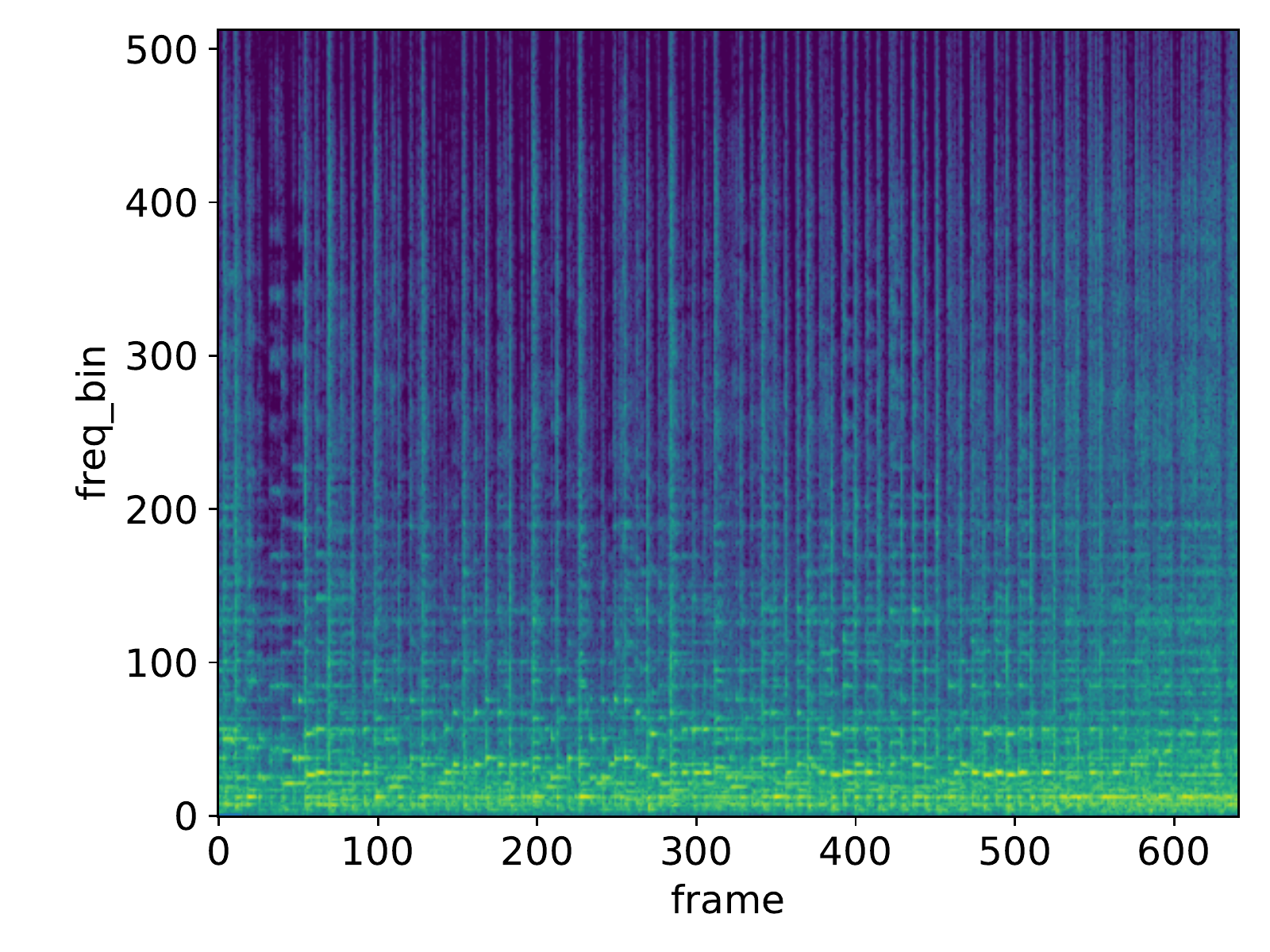}
		\end{minipage}
	}
	\caption{The spectrogram and waveform of generated music examples. The model can generate diverse music, including smoothing and cadenced (a, c) and fast-paced (b, d) rhythms. Text of (a, c): The piano piece is light and comfortable yet deeply affectionate. Text of (b, d): A passionate, fast-paced guitar piece.} \label{fig:spectrogram}
 \vspace{-0.5cm}
\end{figure*}

\subsection{Analysis}
\paragraph{Diversity}
The music generated by our model has a high level of diversity. For melody, our model can generate music with a softer and more soothing rhythm or more passionate and fast-paced music. For emotional expression, some music sound sad, while some are very festive and cheerful. For musical instruments, it can generate music composed by various instruments, including piano, violin, erhu, and guitar. We select two examples with apparent differences and analyze them based on the visualization results. As shown in the waveform from Figure \ref{fig:spectrogram}, the fast-paced guitar piece has denser sound waves, while the piano pieces have a slower, more soothing rhythm. Moreover, the spectrogram shows that the guitar piece holds dense high and low-frequency sounds, while the piano piece is mainly in the bass part.

\begin{figure}[!htb]
	\centering
	\includegraphics[width=.4 \textwidth]{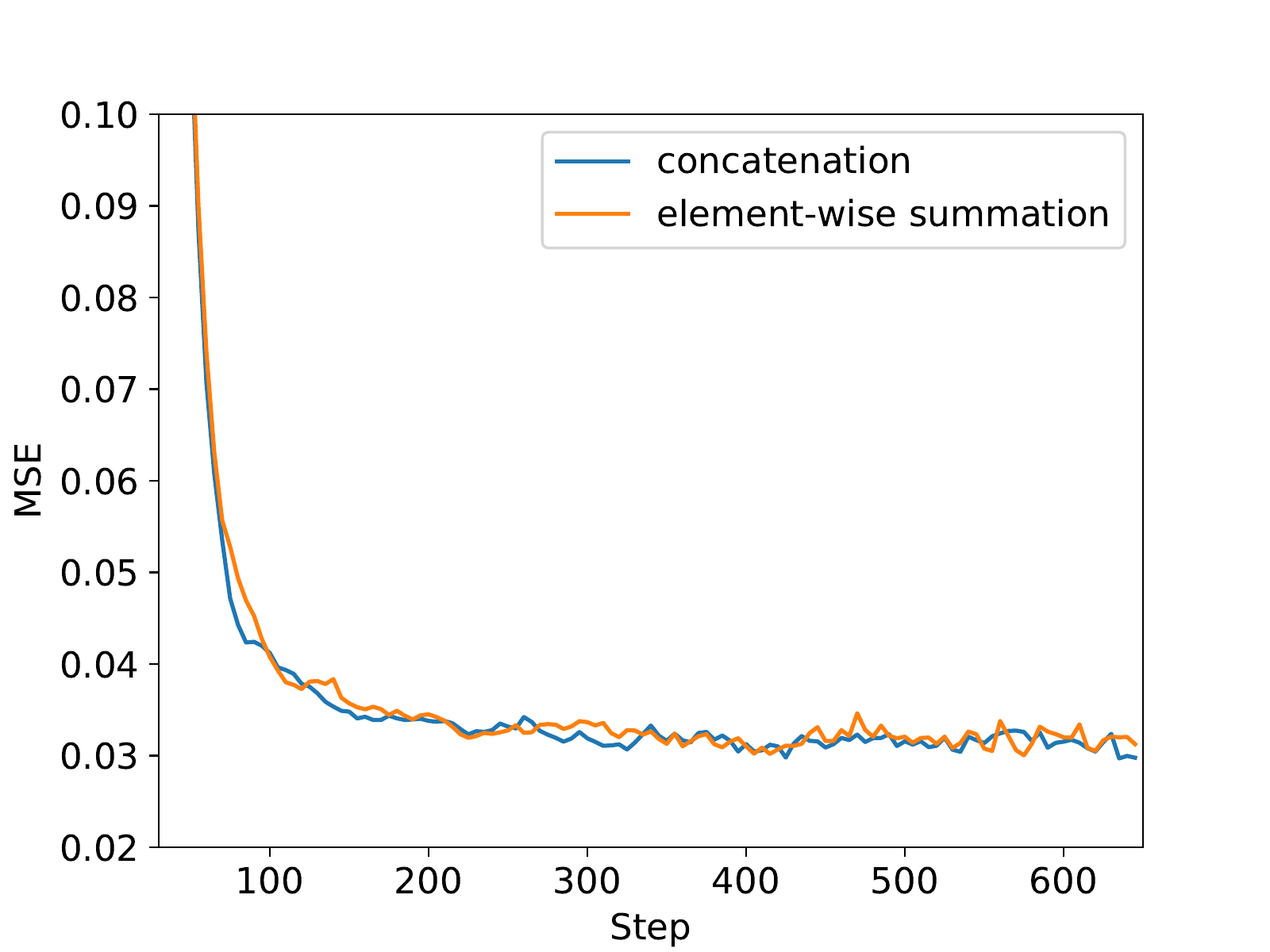}
	\caption{The MSE results on the test set for two implementations of the fusing operation.}
	\label{fig:mse_testset}
 \vspace{-0.5cm}
\end{figure}

\paragraph{Comparison of Different Text Condition Fusing Operations}\label{compare_fuse}
As introduced in Section \ref{timestep_condition}, we compare two implementations of the fusing operation ${\rm Fuse}(\cdot, \cdot)$, namely concatenation and element-wise summation. To evaluate the effect, we compare the performance on the test set as the training progresses. For every 5 training steps, we adopt the model checkpoint to generate pieces of music based on the texts in the test set and calculate the Mean Squared Error (MSE) of generated music and gold music from the test set. The visualization results shown in Figure \ref{fig:mse_testset} indicate no apparent difference between the two fusing operations, thus we adopt the element-wise summation for simplicity.

\paragraph{Comparison of Different Formats of Input Text}
Our proposed method leverages free-form text to generate music. However, considering that the more widely used methods in other works generate music based on a set of pre-defined music tags representing the specific music's feature~\cite{zhang2020butter}, we compare these two methods to obtain better text-music relevance of generated music:
(1) \textit{End-to-End Text Conditioning}.
Suppose the training data consists of multiple text and music pairs $<Y, X>$. The texts in $Y$ are free-form, describing some scenario, emotion, or just a few words about music features. We adopt the straightforward way to process the texts: to input them into the text encoder ${\rm E}(\cdot)$ to obtain the text representations. It relies on the natural high correlation of the $<Y, X>$, and the conditional diffusion model dynamically learns to capture the critical information from the text in the training process. 
(2) \textit{Music Tag Conditioning}.
Using short and precise music tags as the text condition may make it easier for the model to learn the mapping between text and corresponding music. We analyze the text data from the training set and distill critical information from the texts to obtain music tags. Examples as shown in Table \ref{tab:comment_example}. The key features of the music in a piece of long text are limited and can be extracted as music tags. 
We randomly select 50 samples from the test set for manual evaluation. Table \ref{tab:comparison_tag_text} shows the evaluation results of the two conditioning methods, which indicates that our proposed free-form text-based music generation method obtains better text-music relevance than using pre-defined music tags. The main reason might be that the human-made music tag selection rules introduce much noise and result in the loss of some useful information from the original text. Thus it is better to use the \textit{End-to-End Text Conditioning} method for the model to learn to capture useful information dynamically. 






\begin{table}[h]\small
\renewcommand\arraystretch{1.1}
	\centering
	\begin{tabular}{p{3.8cm}|p{0.8cm}|p{0.8cm}|p{0.8cm}}
		\hline
		Method & Score$\uparrow$ & Top Rate$\uparrow$ & Bottom Rate$\downarrow$ \\
        \hline
        Music Tag Conditioning & 1.7 & 22\% & 52\% \\
        \hline
        End-to-End Text Conditioning & \textbf{2.3} & \textbf{40\%} & \textbf{10\%}   \\ 
		\hline
	\end{tabular}
 \caption{\label{tab:comparison_tag_text} Comparison of text-music relevance between two conditioning text formats.}
 \vspace{-0.1cm}
\end{table}

\section{Conclusion}

In this paper, we present ERNIE-Music, a novel music generation model that directly creates music from free-form text. To overcome the scarcity of text-music parallel data, we collect music paired with descriptive comment texts from the internet. We investigate the impact of text format on text-music relevance by comparing two text conditioning methods. Our results showcase ERNIE-Music's ability to generate diverse, coherent music, outperforming existing approaches in music quality and text-music relevance.


\section*{Limitations}

While our model successfully generates coherent and pleasant music, it is important to acknowledge several limitations that can be addressed in future research. The primary limitation is the fixed and relatively short length of the generated music. Due to computational resource constraints, we were unable to train the model on longer sequences. Altering the length during the inference phase can negatively impact performance, which is an area for further investigation.

Another limitation is the relatively slow speed of the generation process. The iterative nature of the generation procedure contributes to this slower speed. Exploring techniques to optimize the generation process and reduce computational overhead could enhance the efficiency of music generation in the future.

Additionally, our current model is designed to generate instrumental music and does not incorporate human voice. This limitation stems from the training data used, which primarily consists of instrumental music. Expanding the training dataset to include vocal music could enable the generation of music with human voice, offering a more comprehensive music generation system.



\bibliography{custom}
\bibliographystyle{acl_natbib}

\appendix
\section{Dataset}
\label{sec:appendix}

Examples of our collected dataset can be seen in Table~\ref{tab:tag_example}.

\begin{table}[h!]\small
\renewcommand\arraystretch{1.5}
	\centering
		\caption{\label{tab:comment_example} Examples of free-form texts and corresponding music tags.}
	\begin{tabular}{p{5cm}|p{2cm}}
		\hline
		Text & Tags \\
        \hline
        {\cn{聆听世界著名的钢琴曲简直是一种身心享受，我非常喜欢}} \newline Listening to the world famous piano music is simply a kind of physical and mental enjoyment, I like it very much & {\cn {钢琴}} \newline piano\\ 
        \hline
        {\cn {钢琴旋律的弦音，轻轻地、温柔地倾诉心中的遐想、心中的爱恋}} \newline The strings of the piano melody, gently and tenderly express the reverie and love in the heart & {\cn {钢琴，轻轻，温柔，爱}} \newline piano, gentle, tender, love\\
        \hline
        {\cn {提琴与钢琴合鸣的方式，在惆怅中吐露出淡淡的温柔气息}} \newline The ensemble of violin and piano reveals a touch of gentleness in melancholy & {\cn {钢琴, 小提琴, 温柔, 惆怅}} \newline piano, violin, gentle, melancholic\\
		\hline
	\end{tabular}
\end{table}

\begin{table}[h]\small
\renewcommand\arraystretch{1.5}
	\centering
            \caption{\label{tab:tag_example} Examples of the adopted and abandoned tags}
	\begin{tabular}{p{1.8cm}|p{4cm}}
		\hline
		 & Tags \\
        \hline
        Adopted & {\cn{希望，生命，钢琴，小提琴，孤独，温柔，幸福，悲伤，游戏，电影}} \newline hope, life, piano, violin, lonely, gentle, happiness, sad, game, movie  \\ 
        \hline
        Abandoned & {\cn{音乐，喜欢，感觉，世界，好听，旋律，永远，音符，演奏，相信}} \newline music, like, feeling, world, good-listening, melody, forever, note, play, believe \\
        \hline
	\end{tabular}
\end{table}

\begin{table*}[!h]\small
\centering
\scalebox{0.87}{
\renewcommand\arraystretch{1.5}
	\begin{tabular}{p{1.6cm}|p{1.5cm}|p{11cm}}
        \hline
        Title & Musician & Text \\
        \hline
        
        {\cn {风的礼物}}\newline Gift of the Wind & {\cn {西村由纪江}} \newline Yukie Nishimura & 
         {\cn{轻快的节奏，恰似都市丽人随风飘过的衣袂。放松的心情，片刻的愉快驱散的是工作的压力和紧张，沉浸其中吧，自己的心。}}\newline The brisk rhythm is like the clothes of urban beauties drifting in the wind. A relaxed mood, a moment of pleasure, dispels the pressure and tension of work. Immerse yourself, your own heart, in it.
     \\
        \hline
        {\cn {九龙水之悦}} \newline Joy of the Kowloon Water & {\cn {李志辉}} \newline Zhihui Li &  {\cn{聆听［九龙水之悦］卸下所有的苦恼，卸下所有的沉重，卸下所有的忧伤，还心灵一份纯净，还人生一份简单。}} \newline Listen to ``The Joy of the Kowloon Water" to remove all the troubles, all the heaviness, and all the sorrows and restore the purity of the soul and the simplicity of life. \\
        \hline
        {\cn {白云}} \newline Nuvole Bianche & {\cn{鲁多维科·伊诺}} \newline Ludovico Einaudi & {\cn{钢琴的更宁静，可大提琴的更多的是悠扬和深沉，也许是不同的演奏方式带来不同的音乐感受吧。}} \newline The piano is more serene, but the cello is more melodious and deep. Perhaps different playing methods bring different musical feelings. \\
		\hline
	\end{tabular}
 }
 \caption{\label{tab:example_data} Examples of our Web Music with Text dataset.}
\end{table*}

\section{Implemention Details}
We train the model for 580,000 steps using Adam optimizers with a learning rate of 4e-5 and a training batch size of 96. We save exponential moving averaged model weights with a decay rate of 0.995, except for the first 25 epochs.

\begin{algorithm}
\caption{Training}\label{alg:train}
\begin{algorithmic}
\Repeat
  \State $x \sim p(x \vert y)$
  \State $t \sim \text{Uniform}([0, 1])$
  \State $\epsilon \sim \mathcal{N}(\mathbf{0}, \mathbf{I})$
  \State $v_t \gets \alpha_t \epsilon - \sigma_t x $ 
  \State Take gradient step on
        \State \hskip 2em  $\nabla_\theta \Vert v_t - \hat{v}_\theta ( \alpha_t x + \sigma_t \epsilon , t, y) \Vert^2 $
\Until{converged}
\end{algorithmic}
\end{algorithm}

\section{Music Tags Extraction}
\label{sec:music_tag}
\noindent To obtain the music tags, we use the TF-IDF model to mine terms with higher frequency and importance from the dataset. Given a set of text $Y$, the basic assumption is that the texts contain various words or phrases related to music features such as instruments and genres. We aim to mine a tag set $T$ from $Y$. We assume two rules to define a good music tag representing typical music features: 1) A certain amount of different music can be described with the tag for the model to learn the ``text(tag)-to-music'' mapping without loss of diversity; 2) A tag is worthless if it appears in the descriptions of too many pieces of music. For example, almost every piece of music can be described as ``good listening"; thus, it should not be adopted as a music tag. Based on such rules, we leverage the TF-IDF model to mine the music tags. Because the language of our dataset is Chinese, we use jieba\footnote{https://github.com/fxsjy/jieba} to cut the sentences into terms. For a term $w$, we make statistics on the total dataset to obtain the TF ${\rm tf}(w)$ and the IDF ${\rm idf}(w)$, then the term score is obtained as ${\rm score}(w) = {\rm tf}(w) \cdot {\rm idf}(w)$. We reversely sort all the terms based on ${\rm score}(w)$ and manually select 100 best music tags to obtain the ultimate music tag set $T$, which can represent the features of music such as instruments, music genres, and expressed emotions. Table \ref{tab:tag_example} displays examples of the adopted and abandoned terms.

\noindent We use the mined music tags to condition the diffusion process. For a piece of music from the training data, we concatenate its corresponding music tags with a separator symbol ``{\cn {，}}'' to obtain a music tag sequence as the conditioning text to train the model.



\end{document}